\documentclass[10pt,conference,final]{IEEEtran}

\pdfoutput=1
\usepackage[utf8]{inputenc}

\usepackage[T1]{fontenc}
\usepackage{url}
\usepackage{flushend}
\usepackage{xspace}
\usepackage[skip=5pt]{caption}
\usepackage[inline]{enumitem}
\usepackage{ifthen}
\usepackage{balance}
\usepackage{hyperref}
\usepackage[numbers]{natbib}
\usepackage{flushend}

\usepackage{microtype}
\usepackage{multirow}
\usepackage[]{siunitx}
\newcommand*\np[2][z]{
\ifx z#1%
$\num{#2}$%
\else%
$\qty{#2}{#1}$%
\fi\xspace%
}

\usepackage{fp}
\usepackage{xfp}

\usepackage{graphicx}
\usepackage{subcaption}
\usepackage{pifont}

\usepackage{booktabs}
\usepackage{multirow}
\usepackage{threeparttable}
\usepackage{xcolor}
\usepackage{array,makecell}
\usepackage{diagbox}
\usepackage{tabularx}

\usepackage{mathtools}
\usepackage{amsmath}

\usepackage{tcolorbox}
\usepackage{pgfplots}
\usepackage{pgfplotstable}

\usepackage{enumitem}
\usepackage[framemethod=TikZ]{mdframed}
\usepackage[scaled]{beramono}

\usepgfplotslibrary{statistics}
\usetikzlibrary{calc,patterns,fpu} 
\pgfplotsset{compat=1.14} 

\pgfplotsset{
    /pgf/declare function={
        Floor(\x) = round(\x-0.49);
    },
    show sum on top/.style={
        /pgfplots/scatter/@post marker code/.append code={%
            \path let \p1=($(normalized axis cs:%
                        \pgfkeysvalueof{/data point/x},%
                        \pgfkeysvalueof{/data point/y})%
                        -(normalized axis cs:\pgfkeysvalueof{/data point/x},0)$)
            in node[
                at={(normalized axis cs:%
                        \pgfkeysvalueof{/data point/x},%
                        \pgfkeysvalueof{/data point/y})%
                },
                anchor={-90*sign(\y1)},yshift={sign(\y1)*2pt}
            ]
            {\pgfmathprintnumber[fixed, precision=1]{\pgfkeysvalueof{/data point/y}}};
        },
    }
}

\definecolor{blau_1a}{RGB}{93,133,195}
\definecolor{blau_2a}{RGB}{0,156,218}
\definecolor{gruen_3a}{RGB}{80,182,149}
\definecolor{gruen_4a}{RGB}{175,204,80}
\definecolor{gruen_5a}{RGB}{221,223,72}
\definecolor{orange_6a}{RGB}{255,224,92}
\definecolor{orange_7a}{RGB}{248,186,60}
\definecolor{rot_8a}{RGB}{238,122,52}
\definecolor{rot_9a}{RGB}{233,80,62}
\definecolor{lila_10a}{RGB}{201,48,142}
\definecolor{lila_11a}{RGB}{128,69,151}

\definecolor{blau_1b}{RGB}{0,90,169}
\definecolor{blau_2b}{RGB}{0,131,204}
\definecolor{gruen_3b}{RGB}{0,157,129}
\definecolor{gruen_4b}{RGB}{153,192,0}
\definecolor{gruen_5b}{RGB}{201,212,0}
\definecolor{orange_6b}{RGB}{253,202,0}
\definecolor{orange_7b}{RGB}{245,163,0}
\definecolor{rot_8b}{RGB}{236,101,0}
\definecolor{rot_9b}{RGB}{230,0,26}
\definecolor{lila_10b}{RGB}{166,0,132}
\definecolor{lila_11b}{RGB}{114,16,133}

\newboolean{showcomments}
\setboolean{showcomments}{true}

\newcommand{\ShowAbsoluteNumber}[1]{%
\ifnum #1<10%
{\hspace*{0pt}#1}%
\else%
\ifnum #1<100%
{\hspace*{0pt}#1}%
\else%
\ifnum #1<1000%
{\hspace*{0pt}#1}%
\else%
{\numprint{#1}}%
\fi%
\fi%
\fi%
}

\newcommand{\ShowPercentage}[2]{%
\FPeval\percentage{round(#1/#2*100,0)}%
\FPeval\percentageOneDecimal{round(#1/#2*100,1)}%
\ifnum \percentage=0%
{\np[\%]{\FPprint{percentageOneDecimal}}}%
\else%
\ifnum \percentage<10%
{\np[\%]{\FPprint{percentageOneDecimal}}}%
\else%
{\np[\%]{\FPprint{percentageOneDecimal}}}%
\fi%
\fi%
\xspace
}
\newlength\BARSIZE  
\newcommand{\inlinechart}[2]{%
\FPeval{\BLACKBARSIZE}{#1/#2}\textcolor{black!80}{\rule{\BLACKBARSIZE\BARSIZE}{1.6ex}}%
\FPeval{\BLACKBARSIZE}{1 - (#1/#2)}\textcolor{black!10}{\rule{\BLACKBARSIZE\BARSIZE}{1.6ex}}%
}

\newcommand*\percent[3][v]{%
\ifx q#1%
    \np{#2}/\np{#3}(\ShowPercentage{#2}{#3})\else%
\ifx s#1%
    \ShowPercentage{#2}{#3}\else%
\ifx p#1%
    \np{#2}(\ShowPercentage{#2}{#3})\else%
\ifx c#1%
    \inlinechart{#2}{#3}%
\else%
    \np{#2}%
    \ifx r#1%
        /\np{#3}%
    \fi%
    \hspace*{0.5ex}(\ShowPercentage{#2}{#3}) %
    \inlinechart{#2}{#3}%
    \xspace
\fi\fi\fi\fi%
}

\makeatletter
\@ifclasswith{IEEEtran}{anonymous}{%
\def\repo{https://anonymous.4open.science/status/EnergyDebug}%
\def\dockerenergy{https://anonymous.4open.science/r/docker-energy}
}{%
\def\repo{https://github.com/enriquebarba97/EnergyDebug}%
\def\dockerenergy{https://github.com/enriquebarba97/docker-energy}
}
\makeatother

\definecolor{mygray}{RGB}{240,240,240}

\newcommand{\answer}[2]{\vspace{.2cm}{\centering\setlength{\fboxrule}{0.1pt}\fbox{\colorbox{mygray}{\parbox{0.95\columnwidth}{\textbf{Answer to RQ#1}. #2}}}\vspace{.2cm}}}

\ifthenelse{\boolean{showcomments}}
 { }
        { }

\definecolor{eminence}{RGB}{108,48,130}
\definecolor{weborange}{RGB}{255,165,0}
\definecolor{frenchplum}{RGB}{129,20,82}
\definecolor{darkgreen}{RGB}{10, 92, 10}

\mdfdefinestyle{mpdframe}{
    nobreak                     =true,
    backgroundcolor             =black!3,
    frametitlebackgroundcolor   =black!15,
    frametitlerule              =false,
    roundcorner                 =5pt,
    innermargin                 =0.3cm,
    innerleftmargin             =0.3cm,
    innerrightmargin            =0.3cm,
    innertopmargin              =0.3cm,
    innerbottommargin           =0.3cm,
    shadowsize                  =1pt
}

\title{Unveiling the Energy Vampires: A Methodology for Debugging Software Energy Consumption}

\author{\IEEEauthorblockN{Enrique Barba Roque}
\IEEEauthorblockA{\textit{Delft University of Technology}\\
\textit{Delft, The Netherlands}\\
enrique@ebarba.com}
\and
\IEEEauthorblockN{Luis Cruz}
\IEEEauthorblockA{\textit{Delft University of Technology}\\
\textit{Delft, The Netherlands}\\
L.Cruz@tudelft.nl}
\and
\IEEEauthorblockN{Thomas Durieux}
\IEEEauthorblockA{\textit{Delft University of Technology}\\
\textit{Delft, The Netherlands}\\
thomas@durieux.me}}

\begin{document}

\maketitle

\begin{abstract}
Energy consumption in software systems is becoming increasingly important, especially in large-scale deployments. However, debugging energy-related issues remains challenging due to the lack of specialized tools. This paper presents an energy debugging methodology for identifying and isolating energy consumption hotspots in software systems. We demonstrate the methodology's effectiveness through a case study of Redis, a popular in-memory database. Our analysis reveals significant energy consumption differences between Alpine and Ubuntu distributions, with Alpine consuming up to 20.2\% more power in certain operations. We trace this difference to the implementation of the \texttt{memcpy} function in different C standard libraries (musl vs. glibc). By isolating and benchmarking \texttt{memcpy}, we confirm it as the primary cause of the energy discrepancy. Our findings highlight the importance of considering energy efficiency in software dependencies and demonstrate the capability to assist developers in identifying and addressing energy-related issues. This work contributes to the growing field of sustainable software engineering by providing a systematic approach to energy debugging and using it to unveil unexpected energy behaviors in Alpine.
\end{abstract}

\section{Introduction}

In recent years, the demand for computing power has grown exponentially, leading to a rapid increase in the number and size of data centers. This growth is accompanied by a significant increase in energy consumption. It is estimated that by 2025, data centers will consume \np[\%]{20} of global electricity and account for \np[\%]{5.5} of global emissions \cite{buyya2024energy}.

While Sustainable Software Engineering and energy efficiency studies have gained traction in mobile development \cite{dornauer2023energy} due to battery life concerns, energy optimization for server deployment remains relatively unexplored. This gap stems from several factors: server systems' lack of reliance on batteries makes energy reduction less immediately impactful, clients do not directly pay for server energy costs, and there's a scarcity of tools for debugging energy consumption in server environments. These circumstances have led to a situation where server-side energy optimization lags behind mobile computing, despite the significant environmental and economic impact of data center energy consumption. Addressing this disparity requires both technological advancements and a shift in perspective regarding the importance of energy efficiency in server-side software engineering.

One of the main components of server software is the Linux distribution over which software runs. In modern server deployments, they are typically bundled with the software into a Docker container, and provide shared libraries over which other technologies run, like the C Standard Library. 

One important criteria for choosing a distribution is image size, which makes images like Alpine extremely popular, with over a billion downloads on DockerHub. However, aspects like energy efficiency are often ignored or unknown by the community.

In this paper, we present a methodology to help developers trace and identify energy consumption hotspots in server systems. Our work is motivated by the findings of \citeauthor{baseimage}~\cite{baseimage}, who demonstrated that the base image of a Dockerfile impacts the energy consumption of the running application. However, the root cause of this energy consumption difference remained an open question.

We introduce a methodology designed to locate the causes of energy consumption discrepancies. This formal approach provides a systematic way for researchers and developers to investigate energy inefficiencies and regressions in workloads that use different libraries or technologies.

To demonstrate the effectiveness of the approach, we present a case study investigating why the Redis database consumes more energy on Alpine than Ubuntu. This study addresses the following research questions:
\begin{itemize}[leftmargin=2.4em]
    \item[\textbf{RQ1}] Does Redis exhibit different energy consumption patterns on different operating systems?
    \item[\textbf{RQ2}] Is our approach capable of identifying the cause of energy consumption differences?
    \item[\textbf{RQ3}] Can the cause for the energy differences be isolated?
\end{itemize}

Our evaluation reveals a difference of \np[\%]{8.6} in total energy consumption and up to \np[\%]{20.2} in power usage during Redis execution on two different operating systems. We attribute this difference to the use of different \texttt{libc} implementations: \texttt{musl} versus \texttt{glibc}. Specifically, we successfully identify the cause of this discrepancy in the \texttt{memcpy} function, which is less performant and more energy-intensive in \texttt{musl}.

In summary, the contributions of this paper are:
\begin{itemize}[leftmargin=1.7em]
    \item A methodology for investigating energy regressions in software
    \item An empirical study highlighting significant energy regressions in the Alpine distribution
    \item A set of scripts and benchmarks for investigating energy regressions in software
\end{itemize}

\section{Background}\label{sec:background}
This section provides essential background information to understand the execution and analysis of our energy experiments. We cover three key areas: containerization technology, C standard library implementations, and energy profiling tools.

\subsection{Containerization and Docker}\label{subsec:docker}
Containerization is a virtualization technology designed to offer an efficient and streamlined approach to software deployment. Unlike traditional virtual machines, containers encapsulate applications and their dependencies, ensuring consistency across different environments and enhancing portability.
Docker, the most popular container platform, facilitates the creation, deployment, and execution of containerized applications. It uses the host machine's kernel, providing a lightweight alternative to full virtualization. This approach allows for seamless movement of applications between development, testing, and production environments, addressing many challenges in modern software deployment pipelines.

\subsection{C Standard Library Implementations}\label{subsec:libc}
The C standard library (\texttt{libc}) API is defined by the International Organization for Standardization, but multiple implementations exist. The most widely used implementation is the GNU C Library (\texttt{glibc}), created in 1987 and part of most Linux distributions, including Ubuntu. Despite its popularity, it has faced criticism for being bloated and slow, with notable figures like Linus Torvalds expressing concerns \cite{mailing}.

An alternative implementation is \texttt{musl} \cite{muslhome}, introduced in 2011. It aims to be lightweight, fast, simple, and standards-compliant, and is the default implementation in Alpine Linux. While \texttt{musl} offers performance benefits, it can face compatibility issues with binaries compiled against \texttt{glibc}, potentially affecting performance in certain scenarios, such as with some Python libraries \cite{pythonAlpine}.

\subsection{Energy Profiling Tools}\label{subsec:energy-profilers}
Energy consumption measurement can be performed using physical power meters or software profilers. Each approach has distinct advantages and limitations:

\subsubsection{Physical Power Meters}
Traditionally used for energy measurement, especially in mobile device studies~\cite{dornauer2023energy}. They provide accurate measurements but may lack granularity in component-level energy consumption.

\subsubsection{Software Profilers}
Modern CPUs allow energy measurements through CPU registers. While less reliable than physical meters, they offer component-level energy consumption data, including individual CPU cores. The most common interface for these measurements is Intel's Running Average Power Limit (RAPL) \cite{rapl}.

Notable software profiling tools include: 
\textbf{\texttt{perf}} \cite{perf}: Measures both computing performance and energy consumption.
\textbf{PowerTOP} \cite{powertop}: Provides detailed power consumption analysis.
\textbf{Powerstat} \cite{powerstat}: Offers power consumption statistics and reporting.

While RAPL is an Intel implementation, AMD provides its own version \cite{Schone_2021} with partial compatibility with Intel's interface. AMD's implementation offers more fine-grained details, such as the energy consumption of individual CPU cores.

\section{Energy Debugging Methodology}\label{sec:methodolgy}

This section presents our main contribution: the energy debugging methodology which simplifies the localization of energy consumption hotspots in software systems. 
This methodology is designed to be a systematic approach that can be applied to any software system that has different versions or configurations that can be tested, and it is the first step in the goal of automating the energy debugging task.
The methodology is divided into four steps: energy measurement, software tracing, aligning trace and energy data, and interpreting results.
To support the methodology, we provide a script to automate the process in our replication package: \url{\repo}.

\subsection{Requirements and framework}

To debug energy consumption using our proposed methodology, a few requirements ought to be considered. 
First, we require at least two versions of the software or system to study. The differences between the two versions help localize potential energy hotspots.
Energy consumption varies depending on the environment and absolute values of energy consumption alone provide meaningless insights. It is therefore required to have a reference software that acts as a baseline to compare the energy consumption.
Second, we require an execution benchmark that performs an execution load on the software. The more representative the benchmark is, the more accurate the results will be. I.e., if the benchmark has a very homogeneous execution, it is unlikely that energy hotspots will stand out.
Finally, we require a server that is ready for reliable energy data collection. We recommend looking at the setup that we use for our evaluation (see \autoref{sec:setup}).

\begin{figure}[t]
    \centering
    \includegraphics[width=0.49\textwidth]{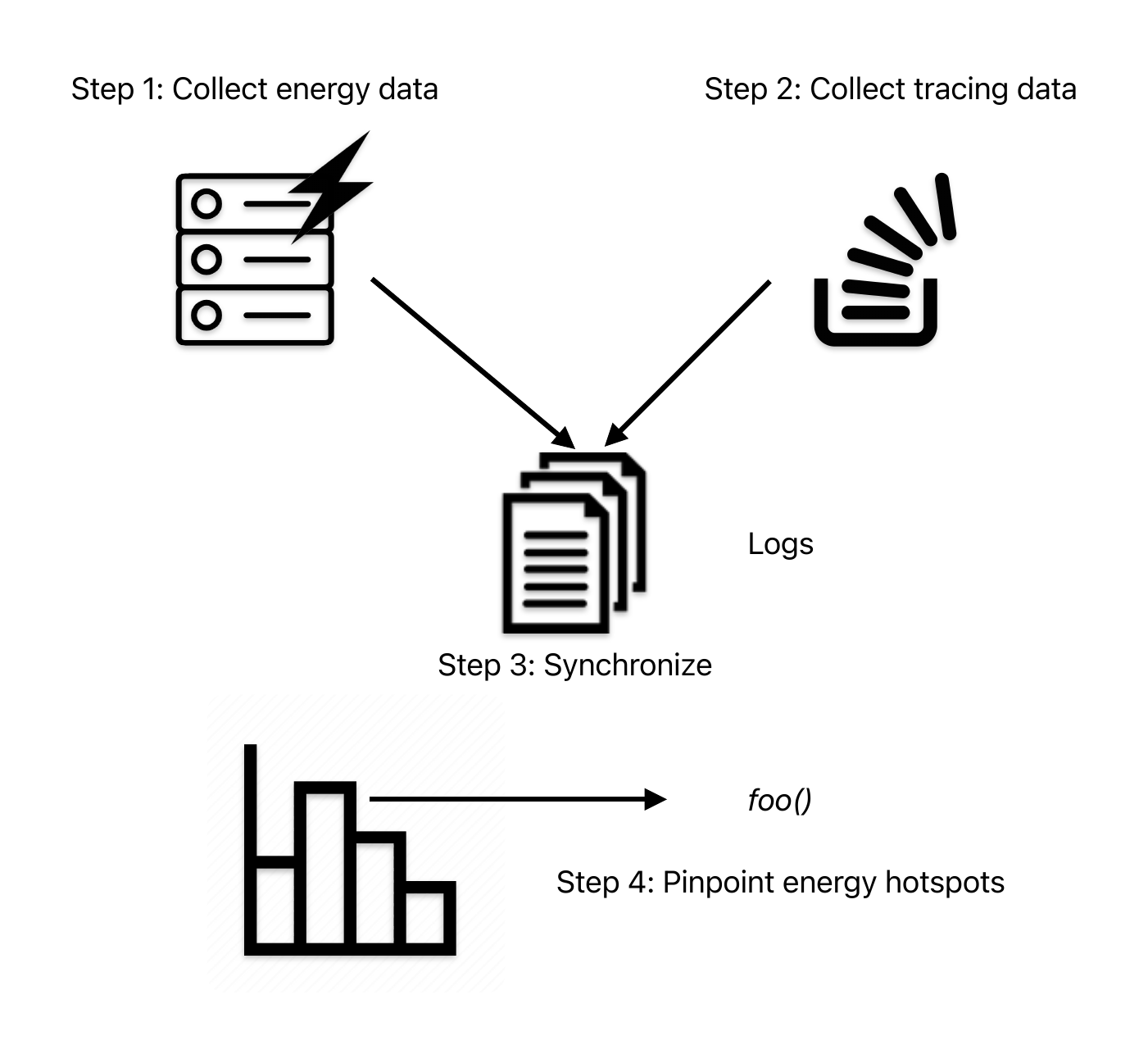}
    \caption{Visual representation of the energy debugging methodology}
    \label{fig:methodology}
\end{figure}

Figure \autoref{fig:methodology} depicts the methodology used for our measurements. It can be summarized in the following steps.

\begin{enumerate}
    \item Run energy benchmarks to identify energy inefficiencies in different sets of dependencies
    \item Run the software with tracing to obtain usage data on the functions from the dependency libraries
    \item Apply log alignment to fix the time dilation introduced by the tracing latency.
    \item Analyze your data. With the normalized data, you can identify the most used functions during the timeframes where higher power usage is observed, which can help identify the function responsible for the difference.
\end{enumerate}

This section details each of these steps, illustrating them by testing a PostgreSQL workload. Section \ref{sec:evaluation} presents the Redis case study and applies the methodology to obtain answers to our research questions.

\subsection{Energy Measurement}

The first step of the methodology is to measure the energy consumption of the different versions of the software.
More specifically, our methodology is designed to highlight the energy consumption differences between two different versions of the software. The difference between these versions do not have to be limited to the software itself, but it can also test different dependencies versions or forks.

We recommend isolating the versions of the software inside a Docker container.
This allows the two versions of the software to run in the same environment while isolating execution to a single core and facilitates switching between versions.
Additionally, it is important that the execution of the software happens in the most systematic way possible.
The executions need to be replicated multiple times and in different order.
We use the framework provided by \citeauthor{baseimage} \cite{baseimage} designed to assist in this process.
It follows the recommended best practices for energy measurements~\cite{energyMeasurementsGuide} such as preload of the system, execution repetition (30x), rest between the executions, randomization of the execution, log management, and energy measurements management.
It also integrates a cross-platform energy measurement tool. 
Additionally, we made small modifications to increase the information and timestamps printed in the logs, to facilitate the log alignment step.
Our fork for this framework is available in our replication package: \url{\dockerenergy}.

\begin{figure}
    \centering
    \includegraphics[width=1\linewidth]{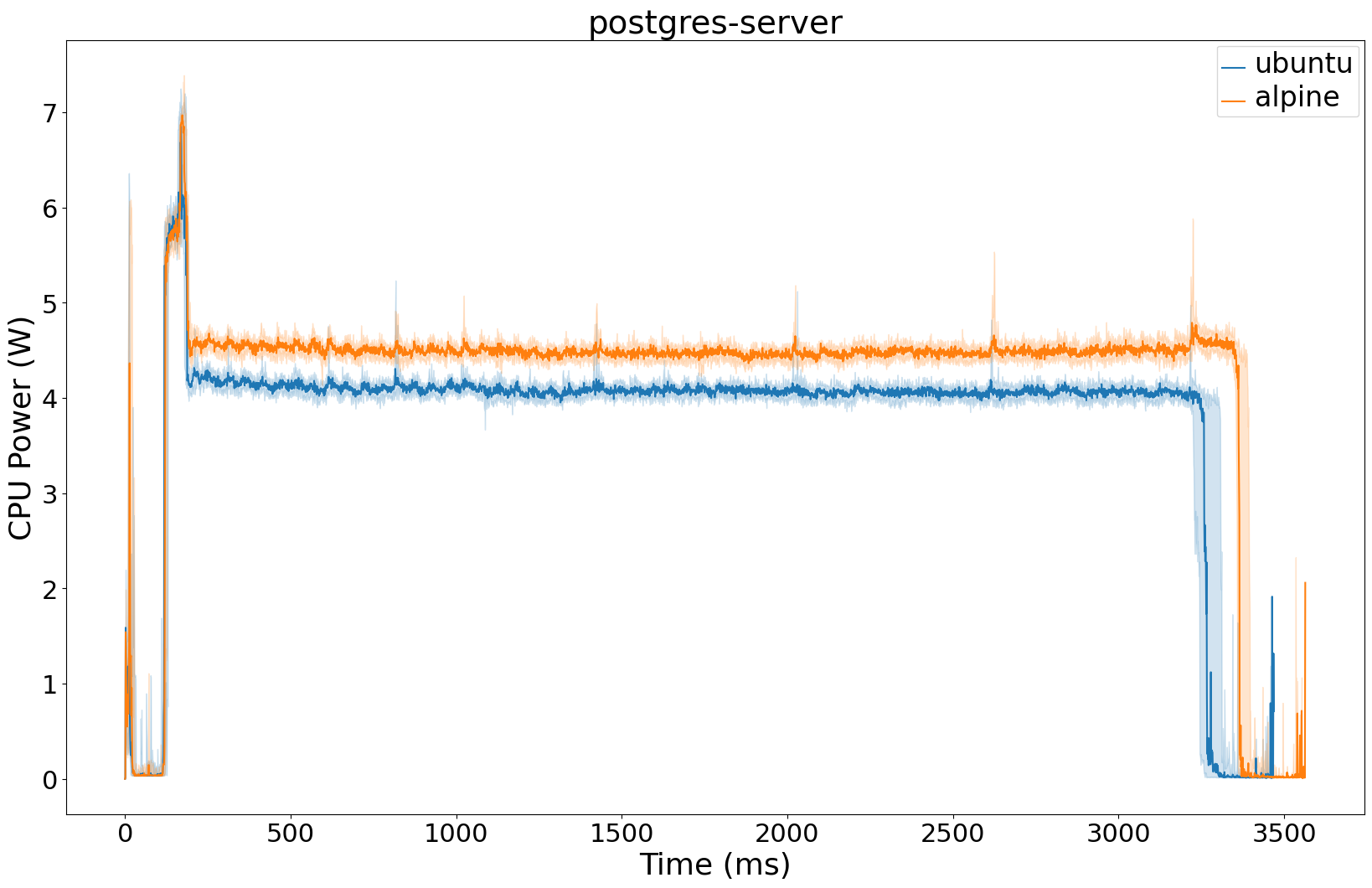}
    \caption{Energy Consumption of PostgreSQL on Alpine vs. Ubuntu.}
    \label{fig:energy_consumption_example}
\end{figure}

At the end of this step, we obtain a list of power consumption over time for every version execution.
\autoref{fig:energy_consumption_example} presents a visual representation of the collected data. In this case, we compare the energy consumption of PostgreSQL on Ubuntu vs. Alpine operating systems.
This figure presents a line with the median power consumption for each operating system and illustrates the variability of the energy consumption with a shading area around the line between the 25th and 75th quartiles. The median energy consumption is the area under the line of the respective operating system.

\subsection{Software Tracing}

In the previous step, we collected the energy consumption. 
However, it indicates a general behavior of energy consumption. 
It is difficult to pinpoint the cause of the energy consumption.
This step aims at collecting software behavior to understand what is happening during the execution.
In this contribution, we collect this software behavior by tracing the function execution of the software.
We use the \texttt{uftrace} tracer \cite{uftrace}. \texttt{uftrace} is an analyzer and tracer for C, C++, Rust, and Python programs. It can track both user space functions and calls to dynamically linked shared libraries. 
The tool reports each call to a function and the duration of said call. 
The program under study normally needs to be compiled using specific options like instrumentation, but the tool also provides the option of dynamically patching certain functions during runtime. 
This patching approach works well for simple shared libraries like the standard library, but it has more problems when trying to trace internal function calls of complex software.

We collect the trace of the software by running the execution benchmark using \texttt{uftrace} once for each version of the software.
We do not need to repeat to step multiple times as this step is (very) time-consuming and produces a (very) large quantity of data. Hence, we separate this execution for trace data collection from the executions for energy data collection.

\begin{figure}
    \centering
    \includegraphics[width=0.49\textwidth]{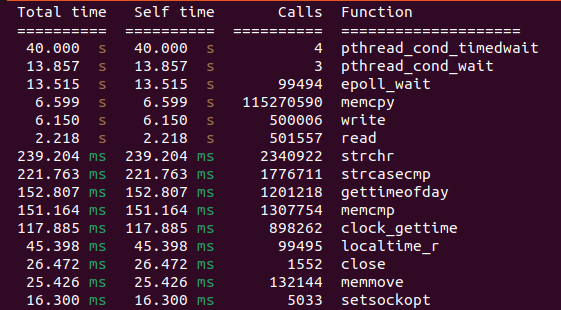}
    \caption{\texttt{uftrace} report of top used functions.}
    \label{fig:uftrace-report}
\end{figure}

\autoref{fig:uftrace-report} presents the trace report of \texttt{uftrace}.
It shows the execution time and the number of calls of each function of the program.
This information is collected by executing the benchmark with the tracing enabled.
This view offers a summary of the execution of the software.
However, akin to our example, if we want to identify the behavior of the software around a specific time this view does not help even if \texttt{uftrace} collects the temporal information.

Indeed, function tracing impact drastically the execution time in a nonuniform manner. It is therefore impossible to directly compare tracing data with energy data.
However, the function tracing can still be used as a large-grain analysis to get the main suspects for the energy hotspot.

\subsection{Pinpoint Energy Hotspots}\label{sec:loc_hotspots}

In the previous steps, we collected the energy consumption and executed functions. 
However, it indicates a general behavior of the software. 
It is difficult to pinpoint the cause of the energy consumption hotspots.
For example, in \autoref{fig:energy_consumption_example} we see that at the beginning of the execution, PostgreSQL consumes the same amount of energy but it changes from timestamp 250, approximately. What happens after 250?

This step aims at providing some suspects. The objective of this idea is to combine energy tracing with function tracers.
The energy tracer aims to measure the energy consumption as accurately as possible, while the function tracer aims to identify what is happening during the execution.

The raw outputs of tracers provide temporary evolution function executions. 
However, an additional problem with this function tracing is that it had a non-uniform overhead during the tracing. This means that not only are the tracers not time-synced, but a linear adjustment for time alignment is also not suitable.
Consequently, it is not possible to align the energy and function tracings directly using the raw data.

To solve this problem, we use the execution log as a reference for time alignment.
For example, if you have a benchmark that executes two tests and at the beginning of each test, the benchmark prints ``[Start] test <X>''.
Those lines can be used as anchors to align the executions.
Our framework of execution prepends the line with the timestamp, it is, therefore, possible to know the execution time of each block on our logs and observe the energy consumption and the executed functions during those moments.  

\begin{figure}
    \centering
    \includegraphics[width=0.49\textwidth]{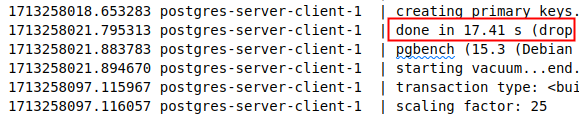}
    \caption{Example of a checkpoint line in PostgreSQL logs.}
    \label{fig:checkpoint}
\end{figure}

This key stoning method uses the execution logs provided to align logically equivalent points. To relate the time points in energy data to time points in tracing data we find checkpoints in the logs. For our study, we define a checkpoint as a relevant and unique or almost unique line in the logs that marks the beginning or end of a section of the workload. In other words, checkpoints are progress marks for the benchmark. An example of a checkpoint from Redis is shown in \autoref{fig:checkpoint}. Redis marks the end of each tested command with some statistics of the execution, preceded by a header that is unique in the logs.

The log alignment process consists of three steps and is designed to work without prior knowledge:
\begin{enumerate}[leftmargin=1.7em]
\item Clean the logs by removing elements that vary over time (e.g., dates, times, IDs, number of requests) to avoid considering periodically reported measurements as unique.
\item Identify relevant lines in the log that can serve as checkpoints. This is an iterative process, starting with unique lines that appear in both logs, then progressively considering lines that appear 2, 3, 4, or more times in both logs until obtaining enough well-distributed checkpoints along the executions. Log lines near the beginning and end of the file are discarded.
\item Align the tracer information by treating time regions between checkpoints as equivalent in terms of execution. The distribution of calls is considered the same in both runs, regardless of timing differences. Function tracing data between each checkpoint is agglomerated and presented as a histogram on the energy measurement tracing graph.
\end{enumerate}

\begin{figure}
    \centering
    \includegraphics[width=0.49\textwidth]{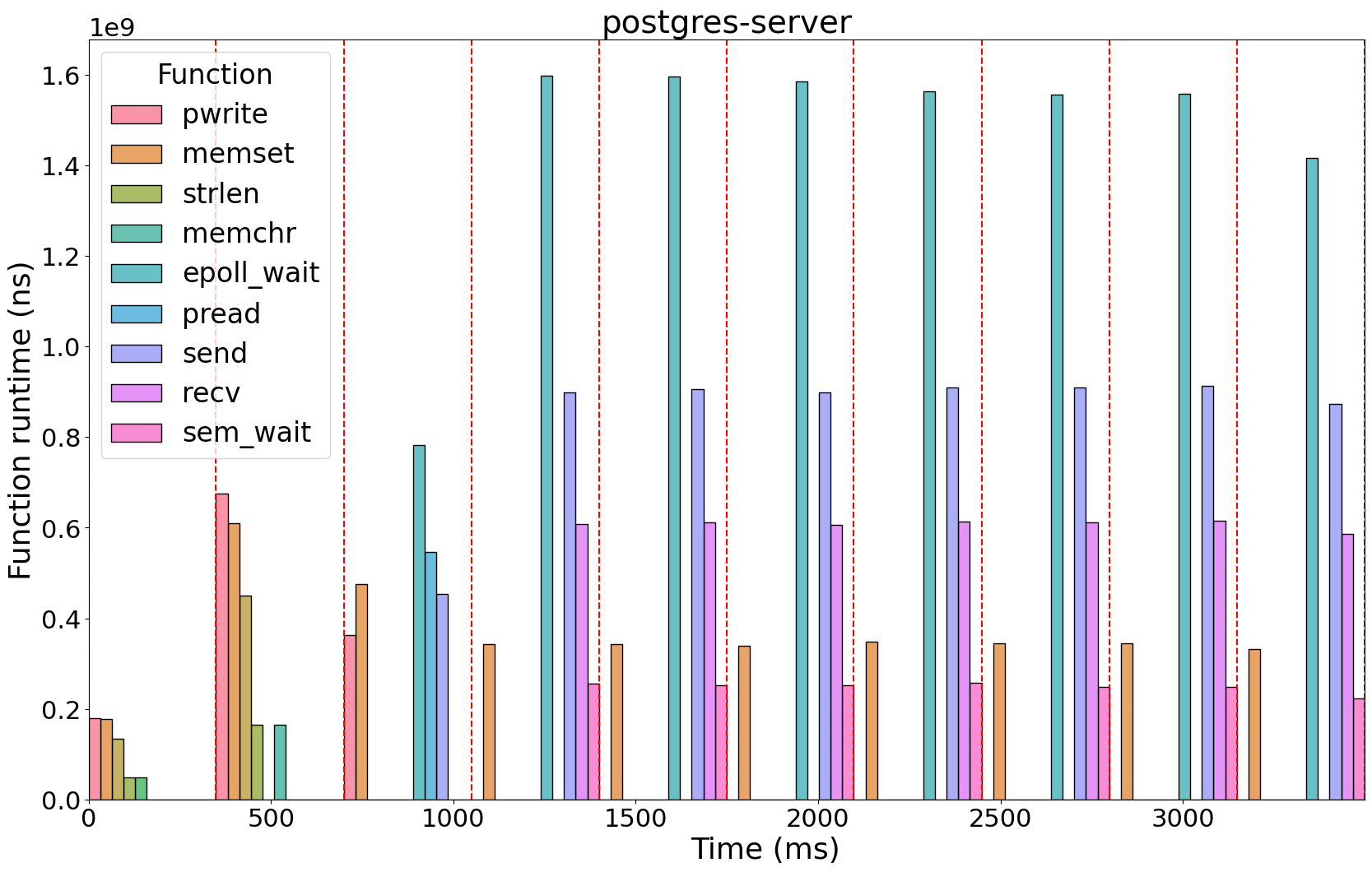}
    \caption{Evolution of the function usage over time with the vertical red lines as the identified checkpoints in the logs.}
    \label{fig:postgres-histogram}
\end{figure}

The result of this step is a graph similar to the one in  \autoref{fig:postgres-histogram}: it shows the evolution over time of the most used functions and it can be compared with the plot of power consumption over time in \autoref{fig:energy_consumption_example} to identify potential energy-draining suspects. The dashed vertical lines depict the checkpoints used in log alignment.

\subsection{Interpretation Results}

The final step of our methodology involves interpreting the tracer results. This step is challenging to automate fully, so we present an example of how to interpret these results.
Comparing Figures~\ref{fig:postgres-histogram} and~\ref{fig:energy_consumption_example}, we observe that the \texttt{pwrite} function dominates the beginning of the execution when there is no difference between Ubuntu and Alpine. However, around the 750-millisecond mark, when the \texttt{epoll\_wait} function starts to appear in the execution, we observe a difference in energy consumption. We also note that the \texttt{send} function is frequent and warrants consideration.
Based on these results, we would investigate \texttt{epoll\_wait} and \texttt{send} by isolating them and performing micro energy benchmarks, as we will present in RQ3 ~\ref{sec:rq3}.

\section{Evaluation}\label{sec:evaluation}

In this section, we evaluate our methodology, by applying it to Redis, a popular in-memory database that is downloaded around \np[M]{60} times a week on DockerHub \cite{redisDocker}.

\begin{itemize}[leftmargin=2.3em]
    \item[RQ1] \textbf{Does Redis exhibit different energy consumption patterns on different operating systems?}
    The first research question aims to confirm the observation from the literature that observed energy consumption can vary depending on the operating system while preserving the performance. We will focus on the Redis case.
    \item[RQ2]\textbf{Is our approach able to locate the cause of the energy consumption difference?}
    The relevance of the second research question is to highlight the performance regression in a system. 
    
    \item[RQ3] \textbf{Can the cause for the energy differences be isolated?}
    In the final research question, we confirm the observation of RQ2 by isolating and testing the identified cause of the energy regression.
\end{itemize}

\subsection{Experimental setup}\label{sec:setup}

\subsubsection{Study Subject}\label{sec:section}
For this evaluation, we chose to evaluate our approach on the Redis database. We chose Redis because it is a popular database that is used by millions of services and has been downloaded more than \np[M]{60} times a week on DockerHub. 
Additionally, Redis was one of the workloads that showed a significant energy usage difference between Alpine and Ubuntu in previous work \cite{baseimage}. 
Moreover, Redis does not have dependencies -- if we compile under the same compiler version and configurations, the energy regression can be isolated in terms of software code being executed.

For our workflow, we use \texttt{redis-benchmark} \cite{redisBenchmark} which is the benchmark provided by Redis.
This benchmark simulates the usage of Redis by $N$ clients to a total of $M$ requests, allowing $N$ and $M$ to be parameterized.

\subsubsection{Tool -  Energy Measurement}

We chose EnergiBridge~\cite{sallou2023energibridge} to measure the energy consumption because it allows us to measure the energy consumption of one specific core on an AMD CPU.
This feature allows the isolation of the measurements and reduces the impact of other running processes. 
Additionally, this tool compared to others allows us to provide energy measurements at regular intervals as well as the CPU, memory usage, and temperature.
EnergiBridge is also compatible with multiplatform (Windows, Linux, OSX) and CPUs (Intel, AMD, Mac ARM).

\subsubsection{Protocol - Energy Measurement}
Measuring the energy consumption of software is difficult because software cannot run in a vacuum, they depend on their environment to be able to be executed: operating system, dependencies, virtualization, memory, hardware.
Additionally, the hardware can also be impacted by external constraints such as the temperature of the room. Consequently, obtaining a precise measurement with a single execution is not guaranteed.

We follow existing guidelines that aim to mitigate those external factors~\cite{energyMeasurementsGuide}.
Before each experiment, we execute a CPU-intensive task for 6 minutes using Sysbench to warm up the CPU and have a consistent temperature.
The next mitigation is to run the energy measurements multiple times in a randomized order, with a short pause between them. The impact of possible variations introduced by unexpected variables, like services in the system, is reduced. 
We perform 30 executions as recommended by existing guidelines~\cite{energyMeasurementsGuide}. 

Additionally, we isolate the workload to a single CPU core and prioritize other processes to run on the other CPU core and measure the energy used by that core.
For example, the energy collection tool will run on a different CPU core and therefore has a limited impact on the energy consumption of the experiment.

Finally, we fixed the CPU frequency and CPU voltage, limited the access to a single user, and limited the number of software installed on the server.

The experiments are run on a server equipped with an AMD Ryzen 9 7900X processor, 64 GB of RAM, and an NVIDIA GeForce RTX 4090 GPU. The server runs Ubuntu 22.04.3, with Linux kernel version 6.2.0, and Docker 24.0.5.

\subsection{RQ1. Redis Energy Consumption}\label{sec:rq1}

As previously mentioned, \citeauthor{baseimage}~\cite{baseimage} shows that different operating systems have an impact on the energy consumption of different software partially in the case of the Redis database.
We identified multiple limitations in the existing study: the version of Redis is not consistent between operating systems, different memory allocators were used, and different compilers.
Those changes could explain the differences. 

To answer RQ1 and confirm the difference in energy consumption between operating systems we replicate the Redis experiments from the previous work~\cite{baseimage}.

\autoref{tab:config} shows the different configurations that are tested. 
We extended the number of cases to standardize the running environment and isolate the potential cause of the energy difference.
For each of our scenarios, we detail the Redis version used, the compiler, and the memory allocator used by Redis.

\begin{table}[t]
\caption{The different setup configurations. The marked cases with a * are the original configuration used in \cite{baseimage}.}
\label{tab:config}
\centering
\begin{tabular}{@{}lllll@{}}
\toprule
\textbf{Label} & \textbf{OS} & \textbf{Compiler}        & \textbf{Redis} & \textbf{Allocator}      \\ \midrule
\texttt{ubuntupack}*                 & Ubuntu & \texttt{glibc} & 6.0.16        & jemalloc       \\
\texttt{alpinepack}*                 & Alpine & \texttt{musl}  & 7.0.15        & musl \\ 
\midrule
\texttt{ubuntulibc}                  & Ubuntu & \texttt{glibc} & 7.2.4         & jemalloc       \\
\texttt{alpineglibc}                 & Alpine & \texttt{glibc} & 7.2.4         & jemalloc       \\
\texttt{alpinejem}                   & Alpine & \texttt{musl}  & 7.2.4         & jemalloc       \\
\texttt{alpinemusl}                  & Alpine & \texttt{musl}  & 7.2.4         & musl \\
\texttt{alpinemusl6}                 & Alpine & \texttt{musl}  & 6.0.16         & musl \\
\bottomrule
\end{tabular}
\end{table}

\begin{figure}[t]
    \centering
    \includegraphics[width=0.45\textwidth]{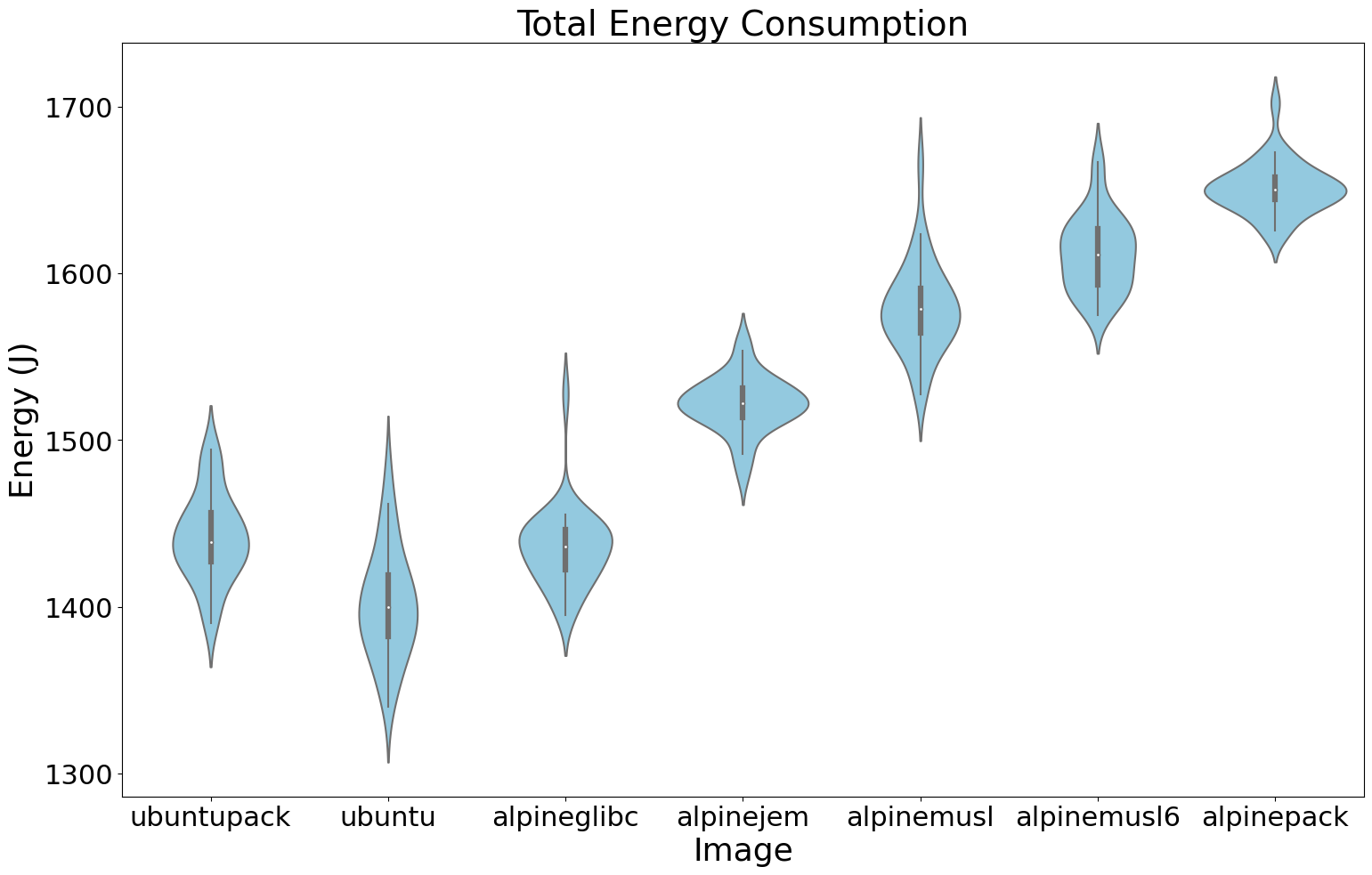}
    \caption{Energy consumption of Redis for the different configurations}
    \label{fig:violins}
\end{figure}

\autoref{fig:violins} shows the violin plots for the total energy consumption of each image and \autoref{tab:consumption} shows the total duration (Time) and average energy usage (Energy). A violin plot is similar to a box plot in the sense that it aggregates the 30 measurements done, showing mean and quartiles. However, instead of having a box shape, it takes the form of the probability distribution of the data. We chose this because it lets us check easily if the experiments are correct. If the shape is symmetrical in the shape of a Gauss bell, we see that the results follow a normal distribution, meaning that the variations in our energy readings are mostly due to random noise rather than systematic errors or unexpected factors.

The first thing we notice in this graph of \autoref{fig:violins} is the energy difference between the images of Redis in Ubuntu and Alpine. The Alpine version uses around \np[\%]{14.5} more energy than the Ubuntu version while taking roughly the same time to complete. These are similar results to the one obtained in~\cite{baseimage}, which validates their experiments.

\begin{table}[t]
\centering
\caption{Average completion time and energy consumption for the different Redis configurations}
\label{tab:consumption}
\begin{tabular}{lrr}
\toprule
\textbf{Image} & \textbf{Time (s)} & \textbf{Energy (J)} \\
\midrule
ubuntupack  & 281.62 & 1441.42 \\
alpinepack  & 286.53 & 1651.11 \\
\midrule
ubuntulibc  & 295.27 & 1401.54 \\
alpineglibc & 284.16 & 1435.28 \\
alpinejem   & 284.15 & 1521.59 \\
alpinemusl  & 288.38 & 1578.99 \\
alpinemusl6 & 286.93 & 1612.10 \\
\bottomrule
\end{tabular}
\end{table}

When fixing and using the latest version of Redis (\texttt{ubuntu} and \texttt{alpinemusl} images) instead of default package manager installations (\texttt{ubuntupack} and \texttt{alpinepack}), the energy gap gets smaller but remains significant. Average consumption improves slightly in Ubuntu between both versions (\np[\%]{2.76} better) and more considerably in Alpine (\np[\%]{6.6} better). This might indicate that there were some performance improvements in Redis between versions. However, there is still a significant gap between Alpine and Ubuntu, with Alpine using around \np[\%]{10} more energy than Ubuntu. This indicates that the discrepancies in Redis versions from the original experiment are not the only reason for the energy performance difference observed.

We can also see how using the custom allocator (\texttt{alpinejem}), which was disabled in the Alpine packaged installation, slightly improves energy usage with respect to the standard library allocator from \texttt{alpinemusl} ($\sim 1\%$ better). However, there is still a significant performance difference of \np[\%]{8.6} between \texttt{alpinejem} and \texttt{ubuntu}, meaning that this configuration difference is also not the main reason for the energy performance difference.

Finally, we look at the energy consumption of Alpine with \texttt{glibc} (\texttt{alpineglibc} image). For this image, results showcase a similar performance between \texttt{alpineglibc} and the Ubuntu images. Except for some outliers, the image performs at around the same level as the old Ubuntu version, with an observed \np[\%]{0.4} improvement, and slightly worse than the most recent Ubuntu, using \np[\%]{2.4} more energy. However, the image performs considerably better than any of the Alpine images with \texttt{musl}. 

\begin{figure}
    \centering
    \includegraphics[width=0.49\textwidth]{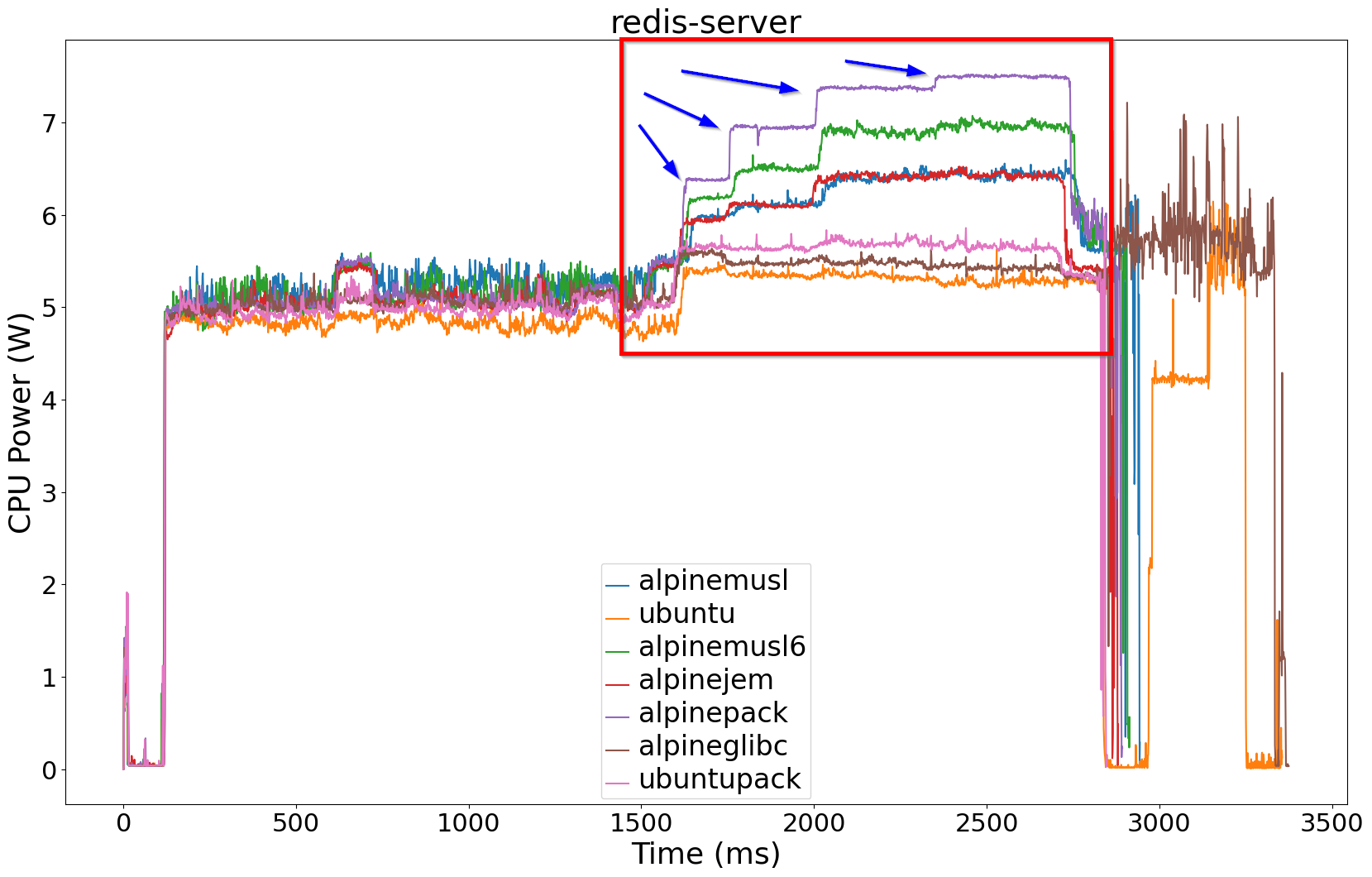}
    \caption{CPU Power usage (W) against runtime of Redis for the different configurations.}
    \label{fig:redis-power-full}
\end{figure}

We further inspect these differences by comparing the power consumption throughout the different executions of the benchmark. \autoref{fig:redis-power-full} shows the median CPU power usage over time for the 30 runs of each image across time during the execution. In this image, we can see that power usage is similar in all experiments for most of the Redis benchmarks, hovering at around 5W. After a certain point, power usage goes up for all images (boxed region). However, this increase is not uniform, and it affects \texttt{musl} images the most. If we compare power usage between \texttt{ubuntu} and \texttt{alpinejem}, images where all configurations are equal except for \texttt{libc}, we obtain a difference of up to $1.17 \textrm{W}$ in the larger gap (around \np[\%]{20.2} difference). This indicates that the main reason for the total energy difference is happening in this part. 

We further investigate this pattern and manually check the logs. We can observe that the operations that are running at this point are LRANGE operations from Redis. The LRANGE command \cite{redisLrange} is a simple instruction that, given a key and a pair of start and end indices, recovers the elements between those two indices. In our benchmark, elements from the database have a default size of 3 bytes, and the LRANGE instruction is tested for 4 different numbers of elements: 100, 300, 500, and 600 elements.

The power usage graph of \autoref{fig:redis-power-full} reveals step increases (highlighted by the blue arrows) that correspond to the increase in the size of the responses from Redis. This correlation, and the fact that energy consumption is similar to the rest of the benchmark, suggests that power consumption depends on workload size. The energy demands of the underlying functions vary linearly with either the argument size or the call number.

\answer{1}{
Alpine and Ubuntu have a substantial difference in power usage and total energy consumption. We confirm that this difference has its origin in the different \texttt{libc} implementations provided by the distributions by introducing \texttt{glibc} into the Alpine distribution. With every other Redis configuration except for \texttt{libc} fixed, we observe a difference of \np[\%]{8.6} in total energy consumption and up to \np[\%]{20.2} in power usage.
}

\subsection{RQ2. Pinpoint the Energy Hotspot}\label{sec:rq2}
Now, we want to find out the reason behind the difference.
To do so, we apply our approach described in \autoref{sec:methodolgy}. 

The first step is to collect the energy measurements. 
We can reuse the measurements from RQ1~\ref{sec:rq1}.
The second step is to collect the execution trace to identify the location of the energy difference.
Tracing is expensive and produces a large amount of data, therefore based on the results of RQ1 we focus the tracing on the \texttt{libc} since it was the main factor of difference. We also adapted the size of the execution, from \np{1000000} to \np{10000} iterations to make the tracing execution manageable.

\autoref{tab:redis-trace-summary} shows the summary of the trace with the top 10 functions that accumulate the longest runtime across the whole benchmark. We see that the most called function \texttt{memcpy}. 
However, the software spends a similar time in the \texttt{write} and \texttt{read} functions, with a much lower number of calls.

\begin{table}[t]
\centering
\caption{Summary of \texttt{libc} function calls and runtimes for Redis}
\label{tab:redis-trace-summary}
\begin{tabular}{@{}lrr@{}}
\toprule
\textbf{Function} & \textbf{Runtime [ns (\%)]} & \textbf{\# Calls [(\%)]} \\
\midrule
\texttt{memcpy} & 6594371865 (\np{39.6}) & 115270590 (\np{92.6})\\
\texttt{write} & 6100079082 (\np{36.6}) & 500006 (\np{0.4}) \\
\texttt{read} & 2204989695 (\np{13.2}) & 501557 (\np{0.4}) \\
\texttt{epoll\_wait} & 701065124 (\np{4.2}) & 99494 (\np{0.1}) \\
\texttt{strchr} & 238221535 (\np{1.4}) & 2340922 (\np{1.9}) \\
\texttt{strcasecmp} & 221697527 (\np{1.3}) & 1776711 (\np{1.4}) \\
\texttt{gettimeofday} & 152283682 (\np{0.9}) & 1201218 (\np{1.0}) \\
\texttt{memcmp} & 150901874 (\np{0.9}) & 1307754 (\np{1.0}) \\
\texttt{clock\_gettime} & 117885050 (\np{0.7}) & 898262 (\np{0.7}) \\
\texttt{localtime\_r} & 45398704 (\np{0.3}) & 99495 (\np{0.1}) \\
\toprule
\end{tabular}
\end{table}

\begin{figure}[t]
    \centering
    \includegraphics[width=0.49\textwidth]{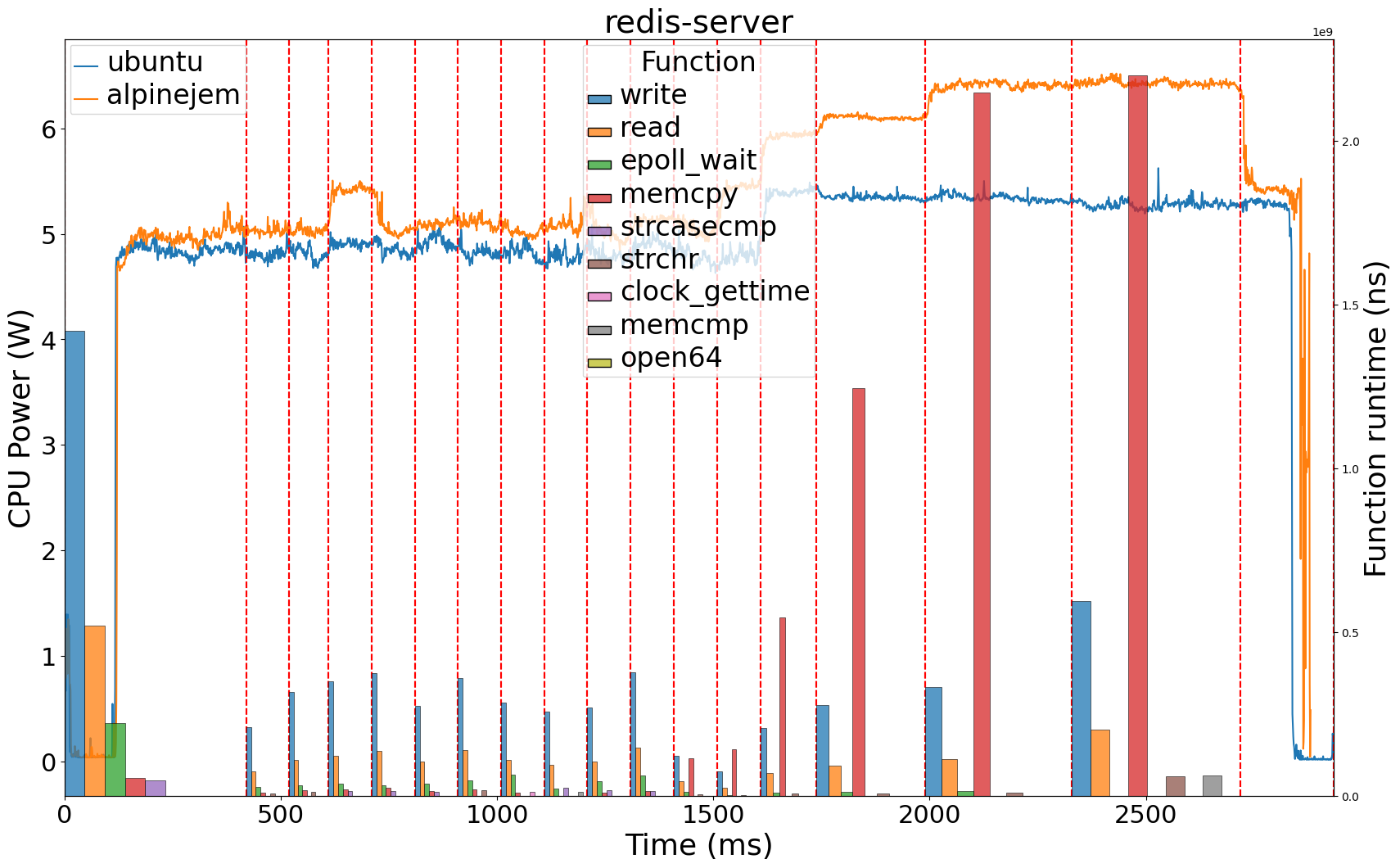}
    \caption{Power usage of Redis for Ubuntu and Alpine and histogram showing function runtime in each region between checkpoints.}
    \label{fig:redis-histogram}
\end{figure}

To gain additional insight into the cause of the energy difference, we perform the third step:  Localize Energy Hotspots (cf. Section~\ref{sec:loc_hotspots}).
This step revolves around applying a technique we coined as \textit{log alignment} to synchronize the trace on the energy graph, allowing us to study the LRANGE portion of the benchmark in detail. 
\autoref{fig:redis-histogram} shows the result of applying this process to the Redis experiment. The dashed vertical lines depict the checkpoints used in log alignment. Between each pair of checkpoints, we draw a bar plot with the distribution of execution time across the most used functions. On top of that, we plot the lines of the power usage of the execution in Alpine (the line at the top, colored in orange) and Ubuntu (the line below, colored in blue).  With this plot, we can easily identify the main functions in the areas of the execution where the difference in energy consumption is more significant (incidentally, within the LRANGE portion of the benchmark). We observe that \texttt{memcpy} is consistently the most used function in those areas. 

Redis uses \texttt{memcpy} to move that element to a memory buffer that is later sent to the client through TCP. Most of the functions tested in the benchmark only recover one element. However, LRANGE has multiple tests that recover up to 600 elements, which is done through an iterator that copies the elements to the buffer one by one. This means that a single LRANGE request has up to 600 more \texttt{memcpy} calls than other commands from the benchmark, increasing the usage of the function.

We performed the same analysis for the PostgreSQL database and identified the \texttt{write} function as the most likely source of the issue. To maintain clarity, we have not included the PostgreSQL analysis in this paper, but the results are available in our replication package (cf. Data Availability section).

Those results seem to indicate a difference in energy performance between \texttt{glibc} and \texttt{musl}.
Interestingly enough, even with an energy difference, it does not seem to have an important performance difference between the two implementations (at least according to this benchmark).
Energy debugging could be used to identify different behaviors between different implementations where execution time does not seem to be identified.

\answer{2}{Our methodology was able to identify the \texttt{memcpy} function from \texttt{musl} as the main suspect for the higher energy consumption for Redis. We also identified the function \texttt{write} as another potential source of energy inefficiencies in PostgreSQL}

\subsection{RQ3. Isolate Energy Hotspot}\label{sec:rq3}

In RQ2~\ref{sec:rq2}, we identify \texttt{memcpy} function as the main suspect.
This research question aims to isolate the difference in energy consumption of \texttt{memcpy} to prove it is the cause of the difference.

To do so, we benchmark the \texttt{memcpy} function and measure the energy consumption. 
The first benchmark consists of a simple \texttt{memcpy} benchmark written in C, adapted from an existing benchmark \cite{memcpyBenchmark}. This benchmark allocates a random memory buffer of 12 GB and performs the following operations:
\begin{enumerate}[leftmargin=1.7em]
    \item Copy all 12 GB to another point in memory with a single \texttt{memcpy} call
    \item Copy all 12 GB with multithreading, to measure multithreaded performance.
    \item Copy all 12 GB in small sequential batches using $2^{20}$ calls to \texttt{memcpy}. This is more similar to the behavior observed in Redis
\end{enumerate}

Each of the operations is repeated 8 times to elongate the duration of the benchmark and be able to take proper energy measurements. Before starting a repetition, the buffers get allocated with \texttt{malloc} and they are freed after each repetition.

\begin{figure}
    \centering
    \includegraphics[width=0.49\textwidth]{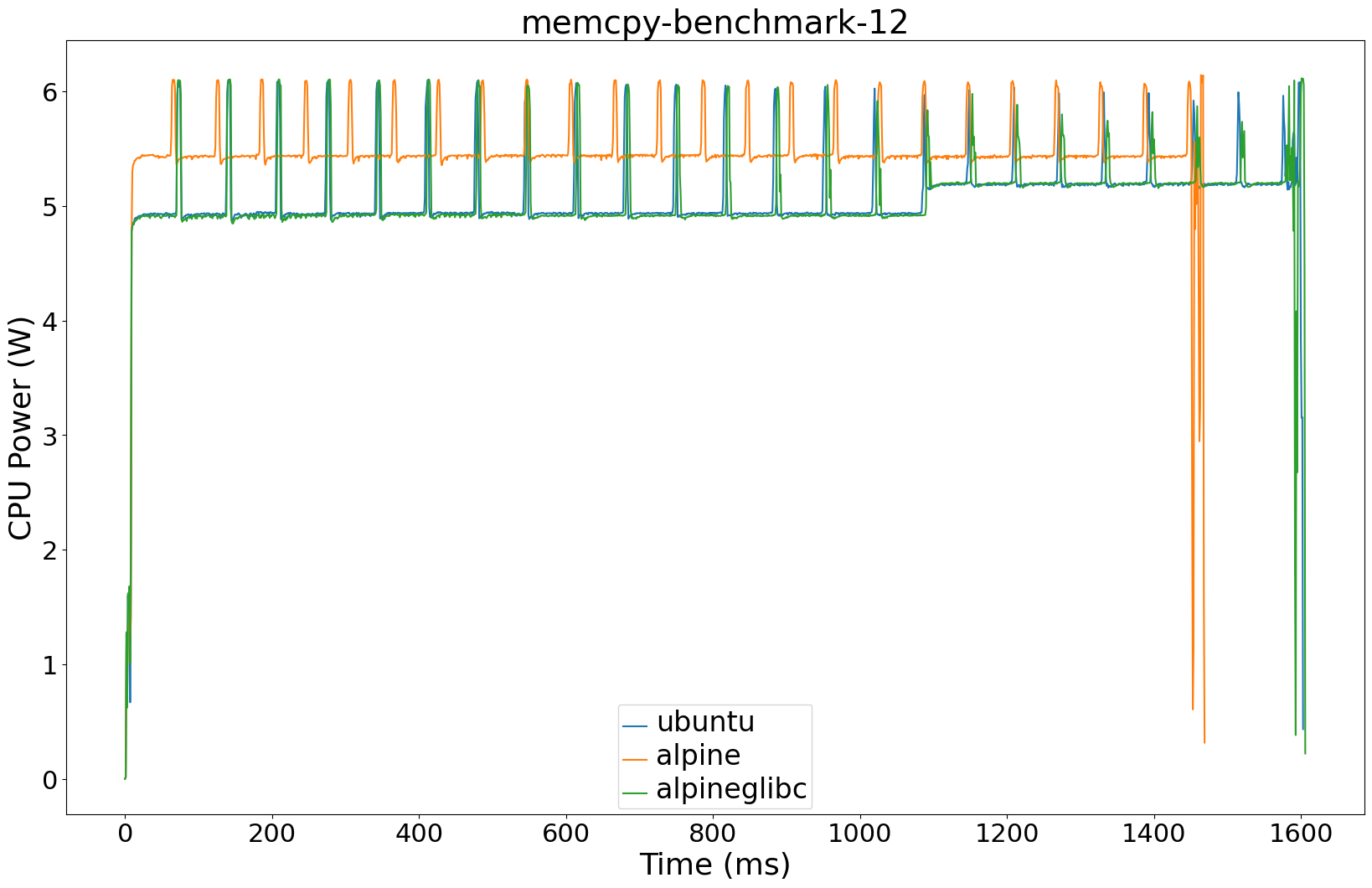}
    \caption{Power usage of the \texttt{memcpy} benchmark with 12 GB.}
    \label{fig:memcpy-benchmark-12}
\end{figure}

\begin{table}[t]
\centering
\caption{Average completion time and energy consumption for the \texttt{memcpy} benchmark with 12 GB}
\label{tab:memcpy-12-consumption}
\begin{tabular}{lrr}
\toprule
\textbf{Image} & \textbf{Time (s)} & \textbf{Energy (J)} \\
\midrule
alpine & 145.39 & 783.48 \\
ubuntu & 158.38 & 795.13 \\
alpineglibc & 158.81 & 793.55 \\
\bottomrule
\end{tabular}
\end{table}

\autoref{fig:memcpy-benchmark-12} shows the power usage of this experiment for Alpine and Ubuntu, and \autoref{tab:memcpy-12-consumption} shows average time and consumption.
 We can observe a slight difference in power usage, with Alpine using slightly more power than Ubuntu, with a difference of around $0.5\text{W}$ or \np[\%]{9.5} for most of the benchmark. At the final step of the benchmark, where we divide the \texttt{memcpy} into $2^20$ sequential calls, energy usage for the \texttt{glibc} based images goes up a bit, while Alpine usage remains the same.

The higher power usage in Alpine is translated into a faster runtime and the Alpine image ends up taking slightly less overall energy to complete the task. 
This result is not the same as the one observed in RQ2. 
The discrepancy arises because this benchmark does not reflect the usage of Redis.
Further analysis reveals that the difference is due to the size of the \texttt{memcopy}.
In the Redis benchmark, only 4 bytes of data are copied, while in the previous benchmark, a big chunk of data is transferred.
This detail is important because of memory alignment. In 64-bit architectures, data in memory is required to be 8-byte aligned~\cite{computerSystems}. This means that moving multiples of 8 bytes is usually less expensive than moving less than 8 bytes since the latter requires additional instructions to guarantee proper alignment.

We designed a new experiment that mimics Redis's usage.
In this experiment, we initialize a memory buffer \texttt{destination} of 3000 bytes and perform \texttt{memcpy} operations with small-sized elements. For each of the sizes defined in the LRANGE benchmark (100, 300, 500, and 600), we copy that number of elements of size 4 bytes and repeat for a large number of times (40 000 000) to simulate the high number of calls in Redis and to have a long enough benchmark so we can measure the energy accurately.

The dummy element we use for this experiment is the literal ``VKX,'', the same dummy data that Redis uses. We provide this data to \texttt{memcpy} in two different ways: in the first experiment, we initialize a second buffer of the same size \texttt{source} with the literal repeated over and over. Then, we call \texttt{memcpy} using two moving pointers, one to \texttt{destination} and another to \texttt{source}. In the second experiment, we use a cached literal and copy it over and over until filling the requested number of elements, with a single moving pointer to the destination buffer. While these two experiments might look functionally the same, they change the data structures used and the information known by the compiler, which changes the version of \texttt{memcpy} that is used for \texttt{glibc}.

\begin{figure}[t]
    \centering
    \includegraphics[width=0.49\textwidth]{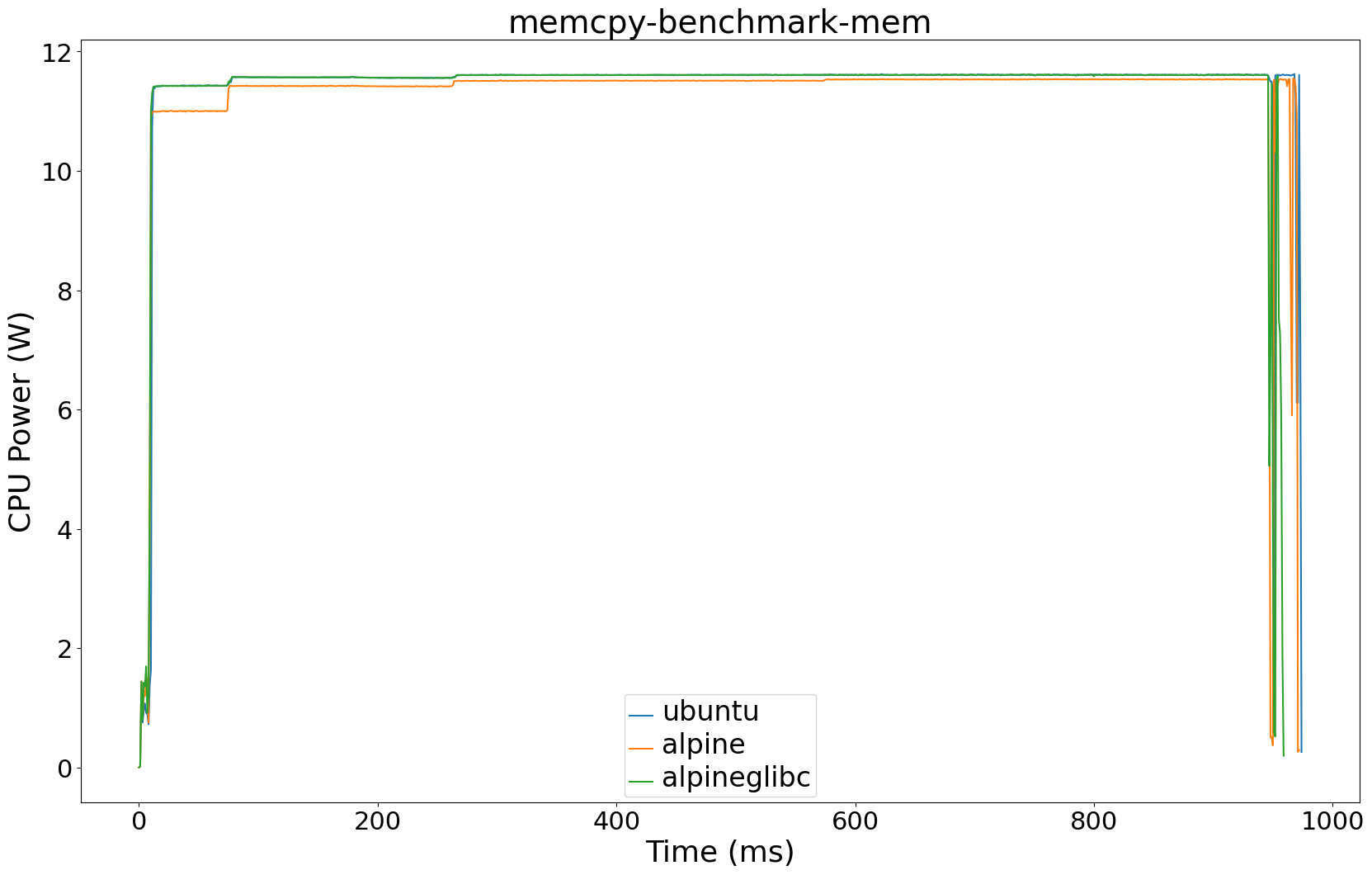}
    \caption{Power usage against time of the \texttt{memcpy} benchmark from memory to memory.}
    \label{fig:memcpy-mem}
\end{figure}

\begin{table}[t]
\centering
\caption{Average completion time and energy consumption for the \texttt{memcpy} benchmark from memory to memory and cache to memory}
\label{tab:memcpy-consumption}
\begin{tabular}{lrr|rr}
\toprule
\multirow{2}{*}{\textbf{Image}} & \multicolumn{2}{c|}{\textbf{Memory to Memory}} & \multicolumn{2}{c}{\textbf{Cache to Memory}} \\\cline{2-5}
 & \textbf{Time (s)} & \textbf{Energy (J)} & \textbf{Time (s)} & \textbf{Energy (J)}\\
\midrule
alpine & 95.05 & 1065.34 & 406.10 & 2977.45 \\
ubuntu & 95.24 & 1075.88 & 155.37 & 972.76\\
alpineglibc & 94.96 & 1075.36 & 142.53 & 869.53\\
\bottomrule
\end{tabular}
\end{table}

\autoref{fig:memcpy-mem} shows power usage over time of the memory-to-memory \texttt{memcpy} experiment, and \autoref{tab:memcpy-consumption} shows average runtimes and energy usage. In this experiment, we observe how the power usage of all images escalates to approximately $11\,\text{W}$. We can also see how, in this case, the \texttt{glibc} images are using more power than \texttt{musl} and, unlike the previous experiment, this is not translated to a shorter runtime. Indeed, now all images take a similar time to complete, and the Alpine image uses $10J$ less to complete the task. We can also notice the small step-ups in power usage as the number of elements to copy goes up, resembling the behavior observed in Redis.

\begin{figure}[t]
    \centering
    \includegraphics[width=0.49\textwidth]{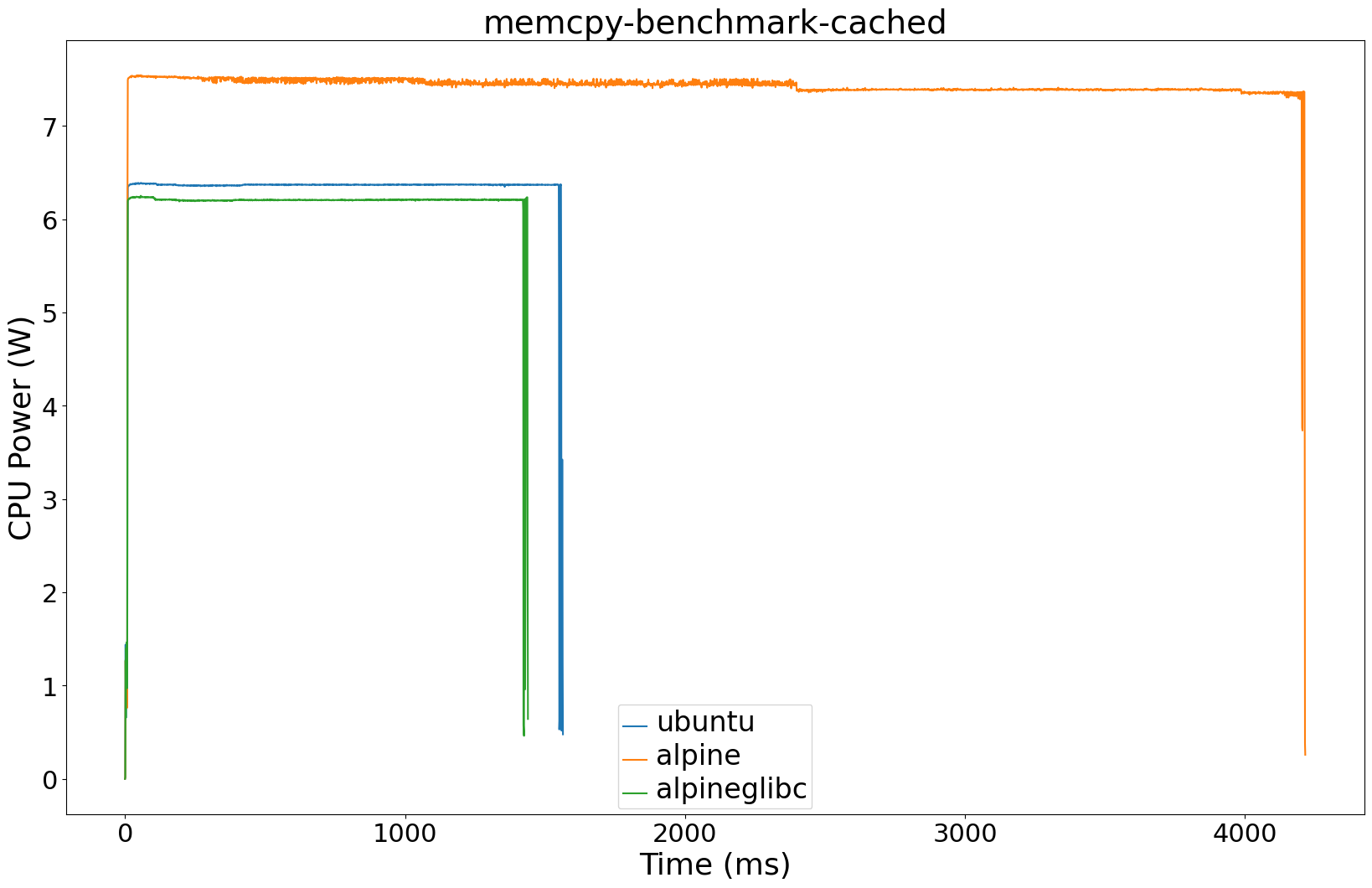}
    \caption{Power usage against of the \texttt{memcpy} benchmark from cached to memory.}
    \label{fig:memcpy-cache}
\end{figure}

\autoref{fig:memcpy-cache} shows the power usage of the cached experiment, and \autoref{tab:memcpy-consumption} compares the runtime and total energy consumption of this experiment. For this benchmark, we obtain completely different results from the previous two benchmarks. Here, \texttt{musl} is much more power hungry than \texttt{glibc}, with a difference of around $1.1W$ or \np[\%]{15.8} difference, closer to the difference observed in Redis. Interestingly, this difference does not translate into more performance, with Alpine taking almost $3\times$ more time to complete the task, resulting in a much higher total consumption.

These benchmarks show that the behavior of \texttt{memcpy} can vary wildly depending on what information is available in compile time. We also notice how some of the benchmarks move in different power usage ranges to Redis. This can be explained because Redis has network communication features. In this benchmark, all \texttt{memcpy} calls are done without pause. However, in Redis, once the \texttt{memcpy} calls for a request are finished, the response has to be sent through TCP, an I/O operation in which the CPU does not have to be used as heavily.

We looked at \texttt{memcpy} implementation in \texttt{glibc} and \texttt{musl}.
\texttt{glibc} seems to have additional optimization with some assembly code~\cite{memcpyGlibc}, while \texttt{musl} does not~\cite{memcpyMusl}. 
\texttt{musl} made the decision to have all its implementation in C however these results showcase an energy hotspot in \texttt{musl} that could be addressed by their developers.

\answer{3}{We successfully isolated the energy difference between operating systems. This confirms that our approach is actually able to locate energy regression. We also notice that the behavior and energy efficiency of \texttt{memcpy} can vary wildly depending on the information available at compile time.}

\section{Discussion}\label{sec:discussion}

\subsection{Complexity of Energy Debugging}
Energy debugging, a subset of runtime analysis, presents unique challenges that surpass those of traditional performance profiling in both complexity and time-intensity. Several factors contribute to this complexity. First, there is a lack of mature profilers. Currently, there are no sophisticated tools to assist developers in identifying energy hotspots effectively. Secondly, a non-linear relationship exists between performance and energy. As our research questions demonstrated, performance and energy consumption do not always correlate directly, which can be counterintuitive. Additionally, while execution time can give hints about the total energy consumption of software, it is not always a good proxy, and direct energy measurements should be used when possible. Finally, modern CPUs introduce a masking effect. The ability of current processors to execute multiple operations per cycle can obscure energy increases without corresponding changes in performance metrics.

Our approach represents an initial step towards simplifying software energy debugging. However, it primarily aids in identifying potential suspects, with manual work still required for confirmation. The design of RQ3 exemplifies this need for manual intervention. Domain knowledge remains crucial for isolating software components that might introduce energy regressions. 

As shown in RQ3, the same logical operations can have different energy impacts in different contexts, which makes energy debugging a difficult task. Energy benchmarks can easily overlook less performant cases, and there is a gap in research on best practices for benchmarking energy efficiency. Additionally, this variety in scenarios makes improving energy efficiency a hand-tailored process that can hardly be generalized to multiple technologies. On top of this, building software with code simplicity and clarity in mind can compromise energy efficiency \cite{energyMaintainability}. This is reflected in the \texttt{memcpy} implementation from \textit{musl}, where the choice for simple code led to energy performance loss in certain scenarios.

As part of our study, we also reached out to the KDE Eco community and presented our results in one of their meetups with senior people from the open-source development community.
\footnote{KDE meetup minutes \url{https://invent.kde.org/teams/eco/opt-green/-/blob/master/community-meetups/2024-06-12_online_community-meetup.md} retrieved on December 10, 2024} 
They confirmed the relevance of our findings and the lack of awareness regarding energy performance in the \texttt{musl} library.

Furthermore, our approach relies on programming language-dependent tracers, highlighting the need for standardization in this area. While we have made progress, there is still a considerable path ahead to streamline energy debugging effectively.

\subsection{Generalization of our Approach}
The complexity of energy consumption debugging is evident in our extensive analysis of Redis and the partial analysis of PostgreSQL. These case studies demonstrate both the potential for generalizing our approach and the need for further automation in the field.

Future research could explore automatic test generation techniques to repeatedly execute suspect functions, potentially automating the isolation of energy regressions. While such automations are beyond the scope of this paper, they represent promising avenues for advancement.

Our primary aim was to demonstrate the feasibility of energy debugging, highlight its challenges, and present a case study that current techniques would struggle to detect and debug. Notably, the authors conducted the debugging of Redis and \texttt{libc} without prior internal knowledge of these systems, underscoring our approach's capacity to assist even those unfamiliar with the target software.


\subsection{Use of Docker for Energy Measurements}

Our evaluation utilized Docker environments to facilitate easy operating system changes through modifications to the Dockerfile base image. While the methodology could be applied to bare-metal systems, containerization simplifies the isolation of multiple instances of the same software and enables easier automation of energy measurements, particularly given the need for repeated executions to ensure accurate observations.

As noted in related work~\ref{sec:related}, Docker introduces only a marginal energy increase compared to bare-metal setups, and this increase remains consistent across executions. Moreover, as mentioned in the background section~\ref{sec:background}, Docker images share the same kernel, thereby reducing variability between executions compared to full virtualization approaches.

\section{Threats to validity}

\subsection{Internal Validity}
Internal validity concerns the extent to which our study design supports the conclusions drawn.

\textit{Representativeness of the benchmark} 
The Redis benchmark used in our study may not fully represent real-world usage patterns and it may not capture all possible Redis use cases. To mitigate this, we used the official Redis benchmark tool, which is designed to simulate typical workloads.

\textit{Measurement accuracy} The accuracy of our energy measurements could be affected by background processes or system noise. To address this, we followed established guidelines for energy measurements~\cite{energyMeasurementsGuide}, including multiple runs, randomized order, and CPU warm-up periods. 


\subsection{External Validity}
External validity concerns the extent to which our findings can be generalized to other contexts.
\textit{Hardware dependence} Our experiments were conducted on a specific hardware configuration (AMD Ryzen 9 7900X processor). The energy consumption patterns and differences observed might vary on different hardware. Future work could replicate the study on diverse hardware to assess the consistency of our findings.

\textit{Operating system versions} 
We tested specific versions of Alpine and Ubuntu. The energy consumption differences might vary with different OS versions or distributions. To partially mitigate this, we tested multiple configurations, including custom-compiled versions of Redis, to isolate the impact of the \texttt{libc} implementation.

\textit{Single core isolation}
To properly isolate the Redis server usage from the client or other processes, we ran the server in a single CPU core following recommended practices. Additionally, our study only measures CPU consumption and not other components like memory or I/O. While it might not reflect the real-life behavior of Redis, our experiments still uncover inefficient behaviors in \textit{musl}'s \texttt{memcpy} implementation that could appear in other applications.

\textit{Software versions} Our study focused on specific versions of Redis and the \texttt{libc} implementations. The energy consumption patterns might change with future software updates.

\subsection{Construct Validity}

Construct validity concerns whether we are measuring what we intend to measure.

\textit{Energy consumption metrics} We primarily used CPU power usage as a proxy for overall energy consumption, which may not capture all aspects of system-wide energy usage. However, given that our focus was on CPU-bound operations like \texttt{memcpy}, this metric is likely to be representative of the energy differences we aimed to study.

\textit{Use of Docker} While Docker introduces minimal overhead and allows for easy comparison between different OS configurations, it may introduce some level of abstraction from bare-metal performance. We mitigated this by ensuring consistent Docker configurations across all test cases and by isolating the workload to a single CPU core.

\section{Related Work}\label{sec:related}

\subsection{Energy Optimization for Docker Images}

Existing tools use AST parsing of Dockerfiles to provide base image suggestions through neural networks \cite{zhang2022recommending} or to detect and warn developers about bad practices \cite{durieux2023parfum}. Other studies propose techniques to assess image quality, size, and build times, based on evolutionary trajectories \cite{Zhang2018evolutionary} or Docker smells \cite{rosa2022smells}. While these studies contribute significantly to their fields, they do not address Docker from an energy perspective.

Several studies compare energy consumption for virtualization and containerization. \citeauthor{morabito2015power}~\cite{morabito2015power} found that for CPU-heavy workloads, power usage is similar across technologies, while container-based technologies show better performance for network tasks. \citeauthor{tadesse2018characterizing}~\cite{tadesse2018characterizing} compared VirtualBox and Docker, confirming similar CPU usage for CPU-intensive tasks, but showed that hypervisor-based virtualization accomplishes less in the same time frame. 
\citeauthor{santos2018does}~\cite{santos2018does} compared Docker's energy efficiency to bare-metal Linux using applications like Redis, PostgreSQL, and WordPress, finding that containers introduce a small, often negligible energy overhead.

In general, Docker demonstrates better energy performance and lower CPU usage compared to virtual machines. However, these studies focus on comparing bare metal, virtual machines, and Docker, without exploring how Docker's energy performance varies with different configurations or base images.

\citeauthor{baseimage}~\cite{baseimage} investigated how base image selection affects Docker containers' energy consumption for various workloads. They tested multiple popular base images like Ubuntu, Debian, and Alpine, providing a framework for running and creating tests. However, their study is limited to reporting energy experiment results without delving into the underlying causes of the observed differences.

\subsection{Energy Efficiency of Languages and Compilers}

\citeauthor{PEREIRA2021102609}~\cite{PEREIRA2021102609} compared the energy efficiency of a wide range of programming languages using diverse programming problem benchmarks. They found that compiled languages like C, C++, and Rust are the most energy-efficient, while interpreted languages like Python or Lua are the least efficient.

For compiled languages, \citeauthor{zambreno2002enhancing}~\cite{zambreno2002enhancing} showed that producing more time-efficient code does not always lead to better energy efficiency, especially for memory performance. \citeauthor{pallister2015identifying}~\cite{pallister2015identifying} studied different combinations of GCC compiler optimization flags and their impact on energy consumption. They found that while compiler optimizations generally improve performance and energy efficiency, some combinations can increase energy usage without improving runtime performance, concluding that no universally optimal set of optimization options exists.

\subsection{Energy Efficiency in Other Software Fields}

Research has identified inefficiencies introduced by commonly used design patterns of reusability and object-oriented programming. \citeauthor{xu2010bloat}~\cite{xu2010bloat} demonstrated how using Java collection objects like \texttt{HashMap} for simple data structures can introduce unnecessary complexity and performance issues. \citeauthor{bhattacharya2011software}~\cite{bhattacharya2011software} highlighted how modern large-scale applications built on deeply-layered frameworks often include unused functionalities that can be an energy burden. Both studies agree that bloated software is a problem exacerbated by diminishing returns from Moore's law \cite{eeckhout2017moore}.

In mobile software development, where battery life optimization is crucial, \citeauthor{dornauer2023energy}~\cite{dornauer2023energy} surveyed elements impacting energy consumption in mobile devices. They found that the main research focus has been on optimizing CPU cycles through efficient scheduling, with hardware improvements for wireless communication also being relevant.

\citeauthor{7972717}~\cite{7972717} studied the effect of mobile software performance best practices on energy consumption. Their experimental study found that following these recommended practices generally leads to improved energy performance. They later expanded this research into a catalog of energy patterns for mobile applications \cite{DBLP:journals/corr/abs-1901-03302}, analyzing commonly applied patterns in Android and iOS applications by collecting commits in open-source projects.

\subsection{Software Energy Profiling}
Other works have previously studied energy usage at a function or library level. \citeauthor{eprof} \cite{eprof} proposed \textit{eprof} as a software profiler that can attribute energy consumption to code locations. However, the energy measurements for the CPU obtained from this tool are based on linear models, which might not accurately reflect the energy usage of current CPUs. Additionally, it requires changes to the OS kernel, which can be difficult to perform for a developer due to technical expertise or lack of privileges over the machine used for experimentation, and these changes might break with newer kernel versions.

\citeauthor{javaenergy} \cite{javaenergy} introduce \textit{JalenUnit}, a unit testing framework to compare the energy efficiency of Java libraries in Java using tracing and instrumentation. While this technique is interesting, the usage is limited to Java workloads and uses statistical sampling to estimate energy consumption, rather than directly measuring energy consumption.

\section{Conclusion}
\label{sec:conclusion}

The increasing demand for computing power, coupled with slowing efficiency gains in hardware \cite{eeckhout2017moore}, has brought software energy efficiency to the forefront of computational challenges. This shift necessitates addressing previously overlooked inefficiencies in software libraries and implementations.

Our research contributes a systematic methodology for identifying energy hotspots in software systems and their dependencies. By effectively pruning libraries under study and pinpointing specific functions or sections with the greatest impact on energy consumption, we significantly narrow the scope of investigation for developers and researchers addressing inefficiencies.

We demonstrated the efficacy of our approach through a case study with Redis, uncovering a notable energy inefficiency in Alpine's \texttt{musl} library. Specifically, we found that certain uses of the \texttt{memcpy} function in \texttt{musl} consume up to \np[\%]{20.2} more power compared to Ubuntu's \texttt{glibc}, with a 13\% difference in a custom benchmark, without runtime improvements. This finding highlights a critical gap in current performance evaluations, which typically focus solely on runtime metrics.

While the Linux and C communities acknowledge performance differences between these libraries, the energy consumption aspect has been largely overlooked. The relevance of these findings where confirmed by experts during our outreach to the KDE Eco Community.

Although the observed difference may seem small for a single instance, its significance amplifies when scaled to thousands or millions of instances across data centers. This research underscores the importance of considering energy efficiency in software development and library choices, particularly in large-scale deployments where cumulative energy savings can have substantial environmental and economic impacts.

\section*{Acknowledgments}
This work was supported by the TU Delft AI Labs program under the SELF Lab.

\bibliography{references}
\end{document}